\documentstyle[psfig,epsfig]{aipproc}
\def\nn{\noindent}

\def\epem{\ifmmode e^+e^-\else $e^+e^-$\fi}
\def\to{\rightarrow}

\def\mpl{\ifmmode \overline M_{Pl}\else $\bar M_{Pl}$\fi}

\begin{document}

\rightline{\vbox{\halign{&#\hfil\cr
SLAC-PUB-8703\cr
November 2000\cr}}}
\vspace{0.8in}

\title{{Randall-Sundrum Phenomenology at Linear Colliders}
\footnote{To appear in the {\it Proceedings of the $5^{th}$ International 
Workshop on Linear Colliders}, Fermilab, 24-28 October 2000}
}

\author{Thomas G Rizzo
\footnote{
E-mail:rizzo@slacvx.slac.stanford.edu. 
Work supported by the Department of Energy, 
Contract DE-AC03-76SF00515}
}

\address{Stanford Linear Accelerator Center\\
Stanford CA 94309, USA}

\maketitle

\begin{abstract}
The physics of the Randall-Sundrum model relevant for future linear colliders 
is briefly summarized. The differences between the case where Standard 
Model(SM) fields are on the wall and where they are in the bulk are emphasized. 
\end{abstract}

Randall and Sundrum(RS){\cite {RS}} have recently proposed a novel approach 
in dealing with the hierarchy problem wherein an exponential warp factor 
arises from a 5-d non-factorizable geometry based on a slice of 
AdS$_5$ space. Here, two 3-branes sit at the orbifold fixed 
points $y=0$ (Plank brane) and $y=\pi r_c$ (SM or TeV brane) with equal 
and opposite 
tensions with the AdS$_5$ space between them. The model contains no large 
parameter hierarchies with $\mpl$, the 5-d Planck scale, $M_5$, and the AdS 
curvature parameter, $k$, being of qualitatively similar magnitudes. TeV 
scales can be generated on the brane at $y=\pi r_c$ if gravity is localized on 
the other brane and $kr_c \simeq 11-12$; indeed in this case the scale of 
physical processes on the SM brane is found to be given by 
$\Lambda_\pi=\mpl e^{-kr_c\pi}$ which is of order a TeV.  

Such a model leads to very interesting and predictive phenomenology that can 
be explored in detail at colliders{\cite {dhr}}. In the simplest scenario the 
SM fields are constrained to lie on the TeV brane while gravitons can 
propagate in the bulk in which case only two 
parameters are necessary to describe the model: $c=k/\mpl$, which is expected 
to be near though somewhat less than unity, and $m_1=kx_1e^{-kr_c\pi}$, 
which is the mass of the first graviton Kaluza-Klein excitation. The masses 
of the higher excitations are given by $m_n=m_1 x_n/x_1$, where the 
$x_n$ are roots of the Bessel function $J_1(x_n)=0$, and are thus not equally 
spaced. While the massless zero mode graviton couples in the usual manner as 
$(\mpl)^{-1}$, the tower states instead couple as $\Lambda_\pi^{-1}$. 

The most distinctive prediction of this scenario is the direct production of 
weak scale graviton resonances at colliders as is shown in Fig.~1 for the case 
of a linear collider. Note that for fixed mass the width of each resonance 
is proportional to $c^2$; for resonances 
beyond the first KK excitation, the width grows as $m^3$. This explains why 
resonances with large KK number tend to get smeared out into a continuum. 
Present searches for graviton resonances at the Tevatron as well as analyses 
of their 
indirect contributions to electroweak observables already place significant 
constraints on the $c-m_1$ plane. When combined with our theoretical 
prejudices the complete allowed region for the RS model is shown in Fig.~1 in 
comparison to the reach of the LHC. Even given some fuzziness in our 
prejudices it is apparent that the LHC should cover the entire RS parameter 
space either by discovering a graviton resonance or excluding the model. 

\begin{figure}
\centerline{
\psfig{figure=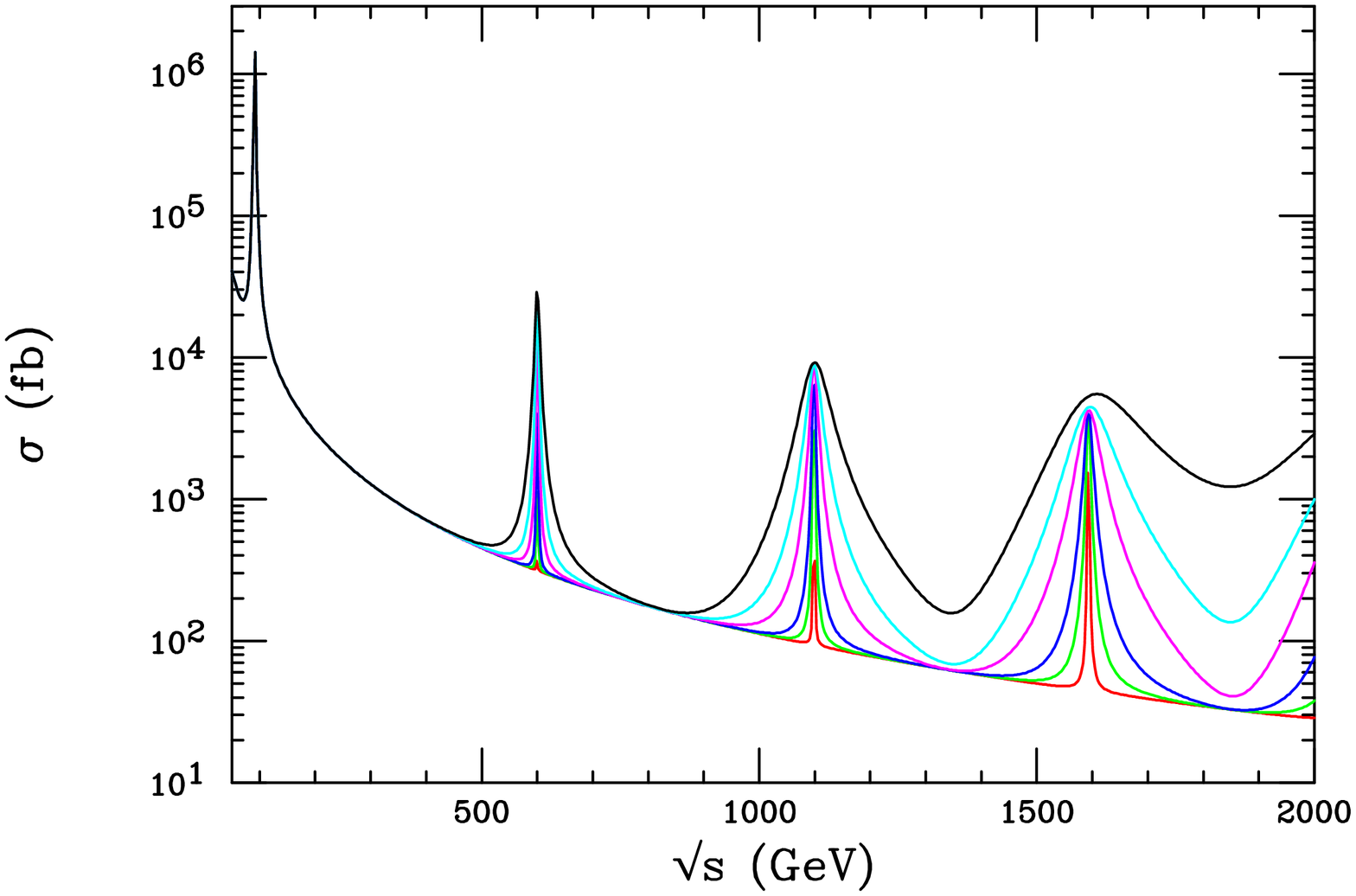,height=7.cm,width=7.0cm,angle=90}
\hspace*{5mm}
\psfig{figure=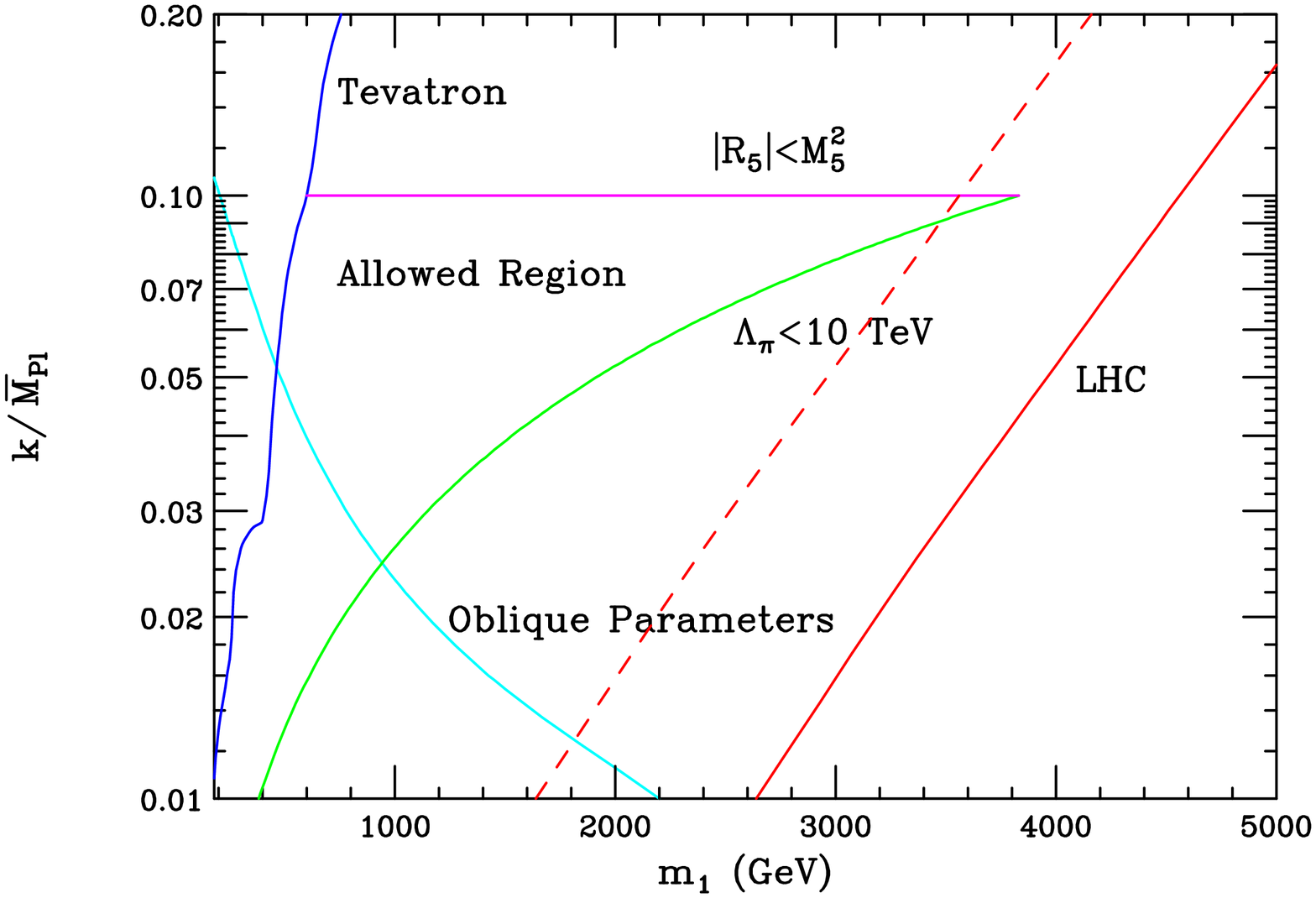,height=7.cm,width=7.0cm,angle=90}}
\vspace*{-0.05cm}
\caption{The left panel shows the production of KK graviton resonances in the 
process $e^+e^-\to \mu^+\mu^-$ assuming $m_1=600$ GeV for various values of 
$c$. The right panel shows the allowed region of the RS parameter space in 
comparison to the discovery region accessible to the LHC with a luminosity 
of 10(100)fb$^{-1}$ 
which lies to the left of the slanting dashed(solid) line. The Tevatron has 
excluded the region to the left of the irregular line on the left whereas an 
analysis of the oblique parameters excludes the region below the smoothly 
falling curve. Our theoretical prejudice that $\Lambda_\pi$ not be too large 
excludes the region below the smoothly rising curve while our similar bias 
that quantum effects do not dominate in the RS setup restricts $c$ to be less 
than $\simeq 0.1$. These last two bounds should be considered `soft' in 
that some fuzziness should be allowed in their direct application as model 
constraints.}
\label{fig1}
\end{figure}

If the SM gauge fields alone are allowed to propagate in the bulk then it can 
be shown that the gauge KK excitations couple much more strongly to the 
remaining wall fields than do the zero modes{\cite {dhr}} by a factor 
$\simeq \sqrt {2\pi k r_c}$. The exchange and mixing of these modes contribute 
to the electroweak observables and result in a bound $\Lambda_\pi > 100$ TeV 
which is perhaps too high to claim a solution to the hierarchy problem. This 
strong bound can be alleviated by also placing the SM fermions in the bulk as 
well with the Higgs field remaining on the wall for a number of technical 
reasons{\cite {dhr}}. 
For simplicity and to avoid FCNC we assume that all SM fermions have an 
identical 5-d mass $m_{5d}=k\nu$, with $\nu$ a parameter of order unity. 
Specifying $\nu$ and $m_1$ for the graviton determines all of the KK masses 
with fermion excitations always more massive than gauge excitations and are 
approximately linear functions of $|\nu+1/2|$. 

%
\nn
\begin{figure}[htbp]
\centerline{
\psfig{figure=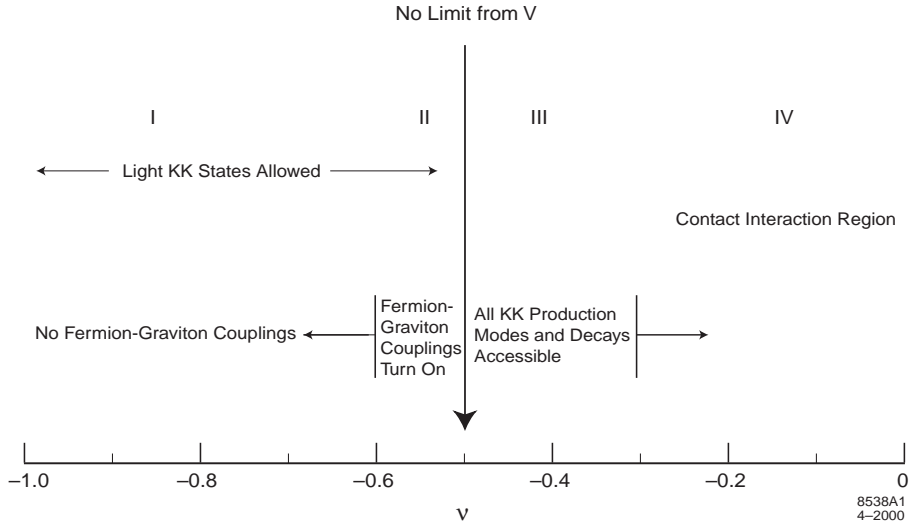,height=7cm,width=12cm,angle=0}
}
\caption{The descriptive phenomenology for each region of $\nu$ as discussed 
in the text. `V' refers to the radiative corrections analysis.}
\label{fig2}
\end{figure}

The phenomenology of this version of the RS 
model is quite $\nu$ dependent as shown in Fig.~2. 
In region IV the masses of the KK states are too large for them 
to be produced at any 
planned collider and exchanges can only be probed via contact interactions as 
shown in Fig.~3. This region is generally disfavoured since it leads to large 
values of $\Lambda_\pi> 10$ TeV. In regions II and III the KK states are 
sufficiently light and their couplings are such that both gauge and graviton 
resonances will be observable at colliders as shown in Fig.~3. In region I, 
the graviton KK states effectively decouple and the gauge KK coupling 
strengths become 
small in comparison to the zero modes. However the gauge KK states will still 
be observable as very narrow excitations at colliders as shown in Fig.~3. 

Clearly future linear colliders can cover all of the possible regions allowed 
when the SM fields are either on or off the wall thus allowing for detailed 
studies of the RS model.

%
\nn
\begin{figure}[htbp]
\centerline{
\psfig{figure=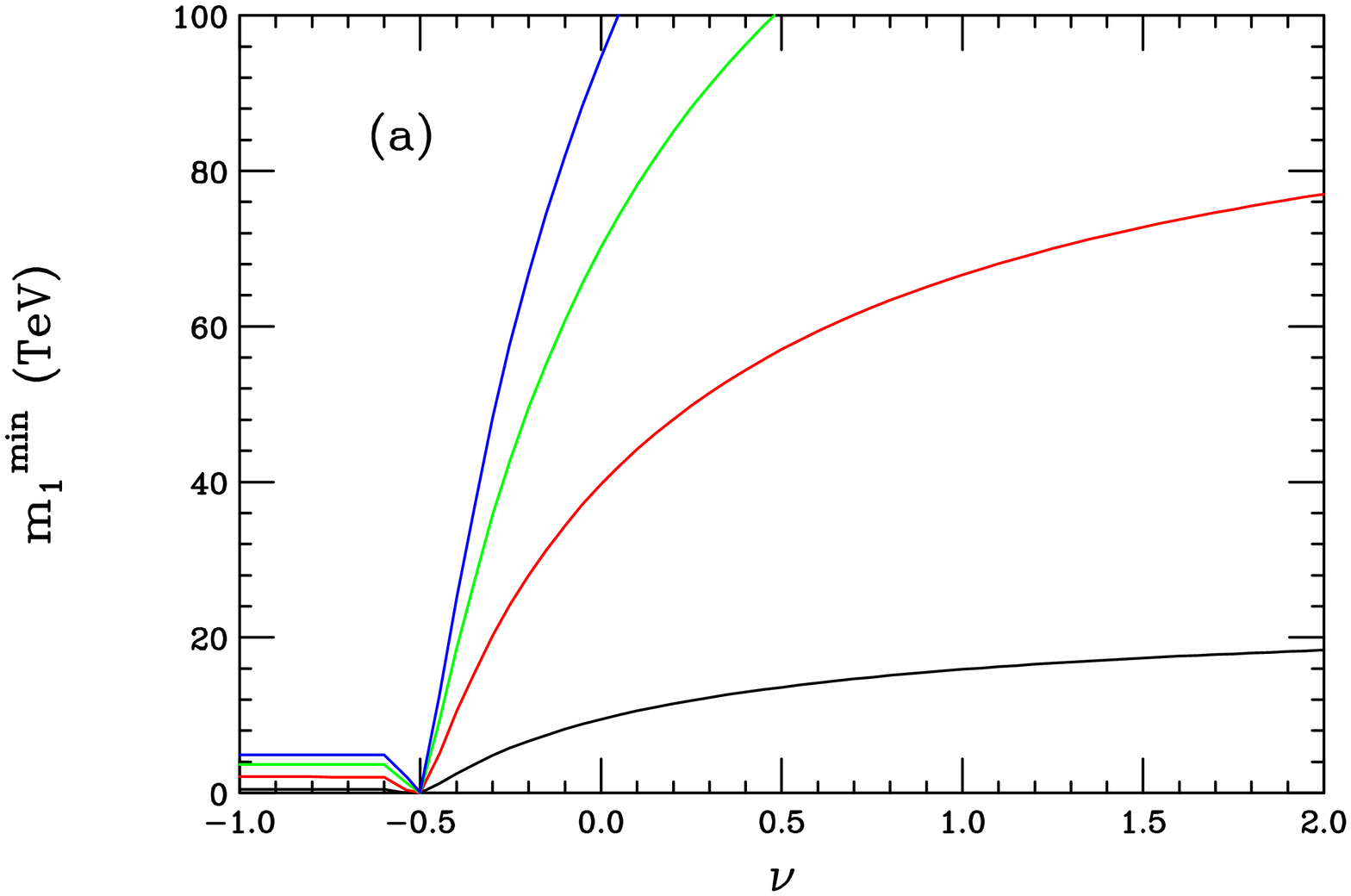,height=7.cm,width=8cm,angle=90}}
\vspace*{0.25cm}
\centerline{
\psfig{figure=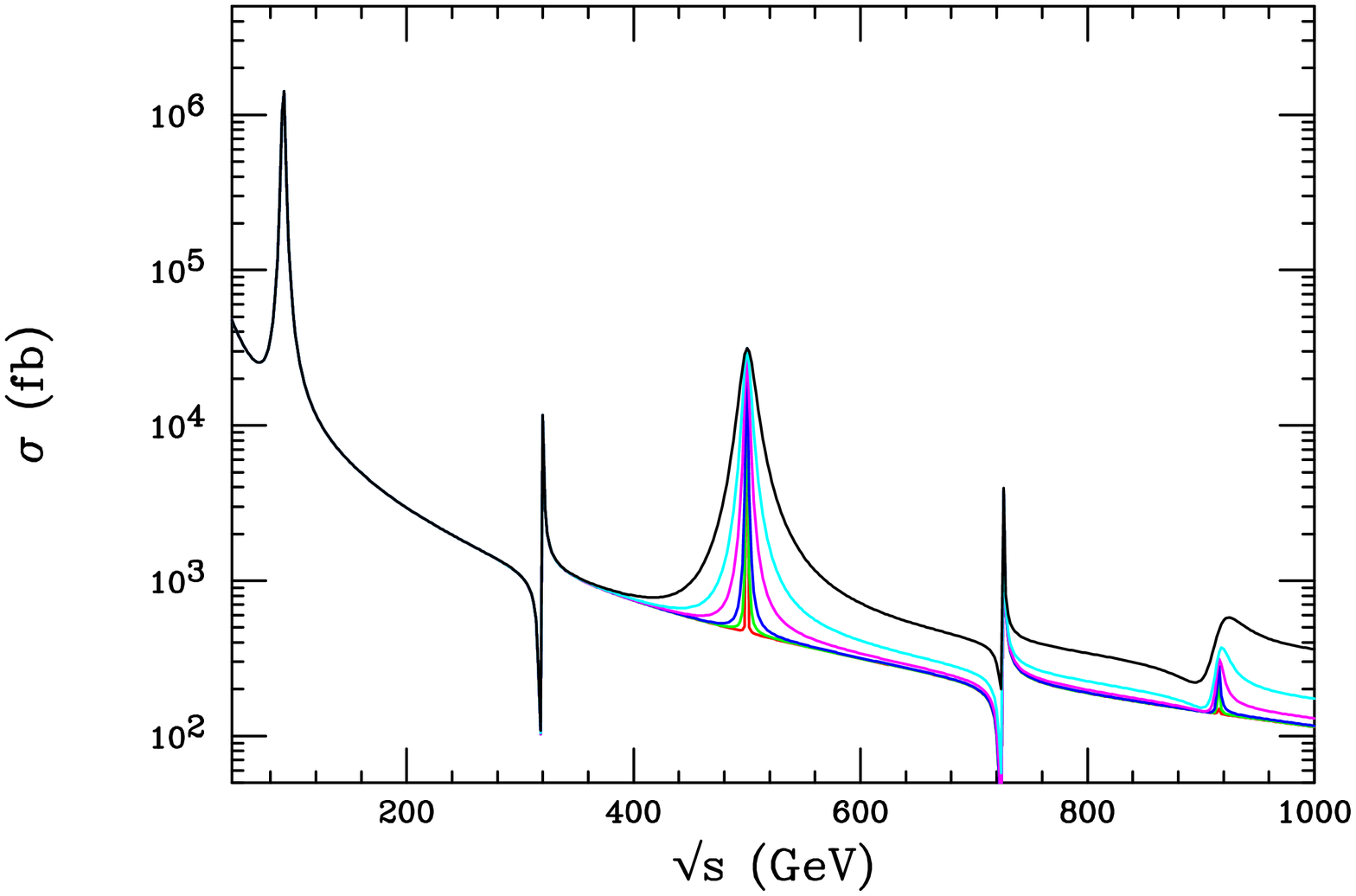,height=7.cm,width=7cm,angle=90}
\psfig{figure=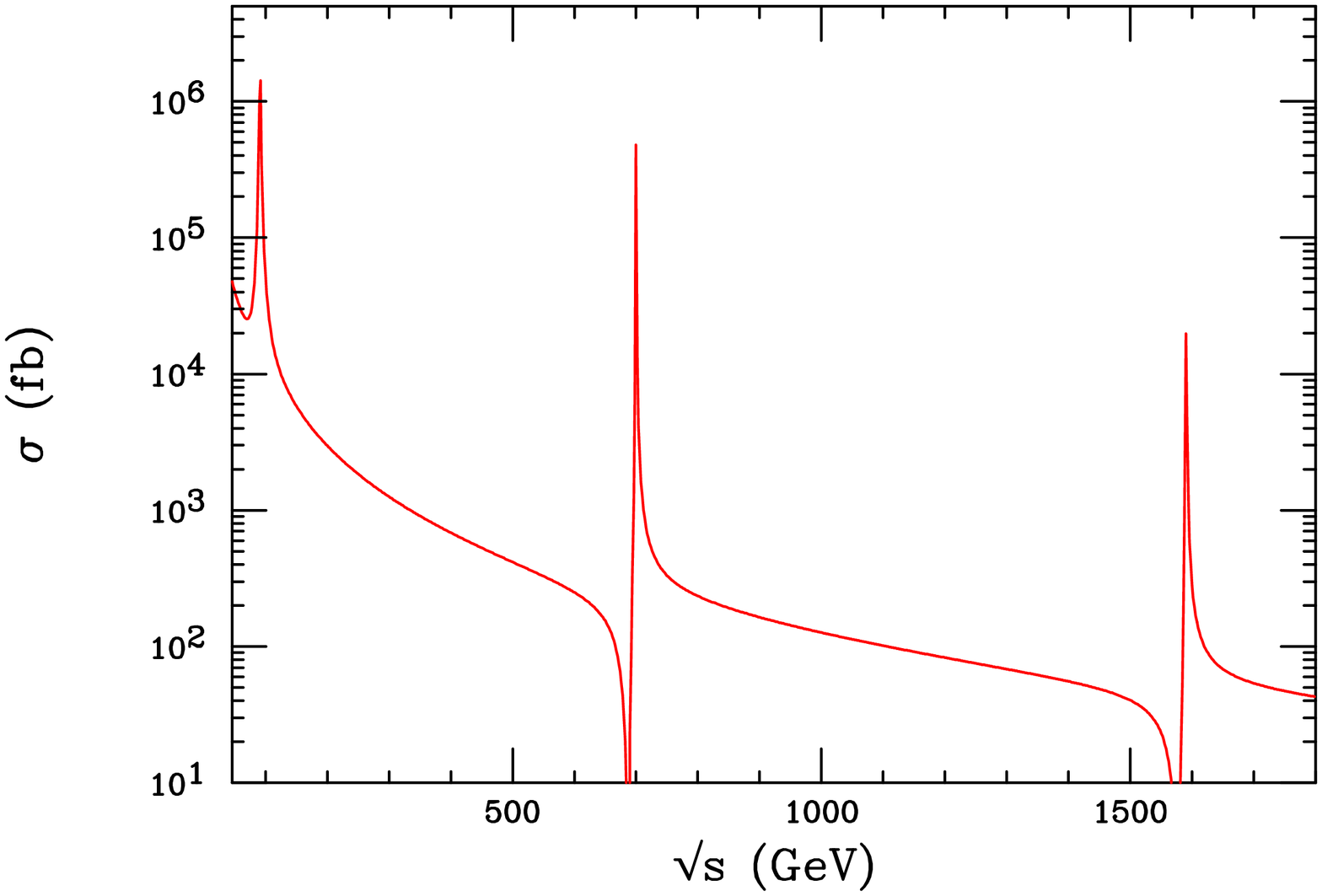,height=7.cm,width=7cm,angle=90}}
\vspace*{0.25cm}
\caption{In the top panel is shown the contact interaction limit on the scale 
of the first KK exchange at lepton colliders. From top to bottom for the NLC 
with 500 fb$^{-1}$ and $\sqrt s=$1.5, 1 and 0.5 TeV and LEP II at $\sqrt s$=195 
GeV with 1 fb$^{-1}$. In the lower left panel one sees the simultaneous 
production of graviton and gauge KK states typical of 
regions II and III via the 
process $e^+e^- \to \mu^+\mu^-$. In the lower right panel for region I only 
the gauge KK states are produced as resonances for this same process.}
\label{fig3}
\end{figure}
%

%
\def\MPL #1 #2 #3 {Mod. Phys. Lett. {\bf#1},\ #2 (#3)}
\def\NPB #1 #2 #3 {Nucl. Phys. {\bf#1},\ #2 (#3)}
\def\PLB #1 #2 #3 {Phys. Lett. {\bf#1},\ #2 (#3)}
\def\PR #1 #2 #3 {Phys. Rep. {\bf#1},\ #2 (#3)}
\def\PRD #1 #2 #3 {Phys. Rev. {\bf#1},\ #2 (#3)}
\def\PRL #1 #2 #3 {Phys. Rev. Lett. {\bf#1},\ #2 (#3)}
\def\RMP #1 #2 #3 {Rev. Mod. Phys. {\bf#1},\ #2 (#3)}
\def\NIM #1 #2 #3 {Nuc. Inst. Meth. {\bf#1},\ #2 (#3)}
\def\ZPC #1 #2 #3 {Z. Phys. {\bf#1},\ #2 (#3)}
\def\EJPC #1 #2 #3 {E. Phys. J. {\bf#1},\ #2 (#3)}
\def\IJMP #1 #2 #3 {Int. J. Mod. Phys. {\bf#1},\ #2 (#3)}


\begin{references}
\bibitem{RS}
L. Randall and R. Sundrum, \PRL 83 3370 1999 .
%
\bibitem{dhr}
H. Davoudiasl, J.L. Hewett and T.G. Rizzo, \PRL 84 2080 2000 ,
\PLB B473 43 2000 ~and hep-ph/0006041.
\end{references}
\end{document}